\documentclass[twocolumn,showpacs,amsmath,amssymb]{revtex4}

\usepackage{epsfig}
\begin{document}
\title{{\em Ab Initio} Structural Energetics of $\beta-$Si$_3$N$_4$ Surfaces}
\author{Juan C. Idrobo, Hakim Iddir, Serdar \"{O}\u{g}\"{u}t}
\affiliation{Department of Physics, University of Illinois at Chicago,
Chicago IL 60607}
\author{Alexander Ziegler, Nigel D. Browning}
\affiliation{Department of Chemical Engineering and Materials Science, 
University of California, Davis, CA 95616, and 
Lawrence Berkeley 
National Laboratory, Berkeley, CA 94720}
\author{R. O. Ritchie}
\affiliation{Department of Materials Science and Engineering,
University of California, Berkeley, CA 94720, and Materials Sciences
Division, Lawrence Berkeley National Laboratory, Berkeley, CA 94720}
\date{\today}

\begin{abstract}

Motivated by recent electron microscopy studies on the Si$_3$N$_4$/rare-earth
oxide interfaces, the atomic and electronic structures of 
bare $\beta-$Si$_3$N$_4$
surfaces are investigated from first principles. The equilibrium
shape of a Si$_3$N$_4$ crystal is found to have a hexagonal 
cross section and a faceted dome-like base in agreement with experimental 
observations. The large atomic relaxations on the prismatic planes are 
driven by the tendency of Si to saturate its dangling bonds, which gives rise
to resonant-bond configurations or planar
$sp^2$-type bonding. 
We predict three bare surfaces with
lower energies than the open-ring $(10\overline{1}0)$ surface
observed at the interface, which indicate
that non-stoichiometry and
the presence of the rare-earth oxide play crucial roles in
determining the termination of the Si$_3$N$_4$ matrix grains.

\end{abstract}  
\maketitle

The desirable mechanical and physical properties of silicon nitride 
ceramics \cite{hoffmann} 
in many high temperature applications are hindered by their 
intrinsic brittleness which limits their wide-spread use and reliability 
as structural components. It has been empirically known for some time
that this problem can be overcome by microstructural and compositional 
design with sintering additives, in particular
rare-earth oxides \cite{sunsat}. The resulting
ceramic microstructure consists of elongated Si$_3$N$_4$ matrix-grains
embedded in an intergranular, typically amorphous, rare-earth oxide phase. 
However, precise information about the 
structure and chemistry of the interface
has been lacking for many years. Recently, there have been three
experimental studies using scanning transmission electron microscopy (STEM), 
which revealed for the first time important information about
the atomic structures and bonding characteristics at the
$\beta-$Si$_3$N$_4$/rare-earth oxide interfaces \cite{shiziegwink}.
This exciting development provides a timely motivation for systematic
theoretical studies of the interface, which should 
complement and aid in the interpretation of these experiments.

Existing theoretical calculations in this field \cite{theory}, which have
typically employed tight-binding, pair potential, 
and molecular orbital methods
(a few of which are from first principles), have mainly focused on the 
bonding sites of additive atoms at the interface and 
the resulting electronic structure. While these studies have provided
important information on the Si$_3$N$_4$/rare-earth oxide interface, 
our approach is to start with detailed first principles calculations
on {\em bare} Si$_3$N$_4$ surfaces, which have not yet been performed. 
We believe that such an approach is the first natural step 
in a {\em systematic} understanding
of the interface at the microscopic level. As it has been historically
the case in {\em ab initio} investigations of 
several other important interfaces, the insights gained
from this step can not only be used as a basis to elucidate the 
bonding characteristics of the additives, but also raise new questions
which need to be addressed to have an in-depth understanding of
the interface.
In this Letter, we present, for the first time, results
from first principles calculations on low-index $\beta-$Si$_3$N$_4$
surfaces focusing in particular on 
(i) the equilibrium shape of the crystal, and 
(ii) the atomic structure and stoichiometry at the prismatic plane 
$(10\overline{1}0)$ surface, 
which is the relevant surface studied in recent STEM experiments [Fig. 1(a)].

\begin{figure}
\includegraphics[width=0.5\textwidth]{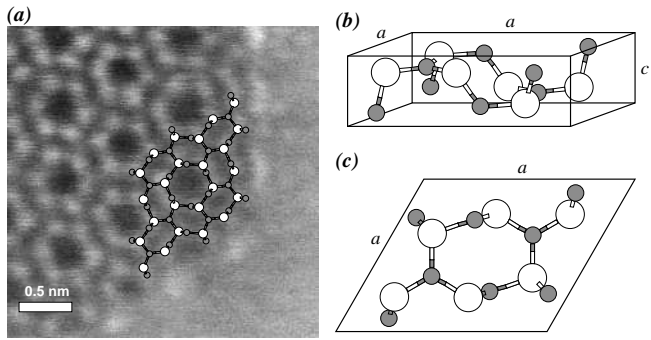}
\caption{(a) Z-contrast image of a Si$_3$N$_4$ grain with a hexagonal edge 
showing the interface with Lu$_2$O$_3$. The grain is oriented with the 
$[0001]$ projection (note the superimposed atomic structure), 
so that the $(10\overline{1}0)$ prismatic boundary 
planes are set at an ``edge-on" condition. Lu atoms are the bright spots
attached in pairs at the termination of the hexagonal open rings. (b)
A perspective view of the $\beta-$Si$_3$N$_4$ unit cell. Si and N 
atoms are shown in white and gray circles, respectively. (c) $[0001]$
projected view of the unit cell.}
\end{figure}

Our calculations were performed within density functional theory
using the projector augmented wave method \cite{vasppaw}. For 
exchange-correlation, we used the Perdew-Wang
parametrization of the generalized gradient approximation (GGA), and
repeated some of the calculations using
the Ceperley-Alder functional within 
the local density approximation (LDA) for comparison.
An energy cutoff of 270 eV was used in all calculations. For 
structural optimization of the bulk, a $\Gamma$-centered
3$\times$3$\times$8 {\bf k}-point grid was employed. Doubling
the {\bf k}-point grid and increasing the cutoff to 400 eV had no
appreciable effect on the calculated structural parameters. 
The hexagonal unit cell of $\beta$-Si$_3$N$_4$ contains
14 atoms with 6 structural parameters 
($a$, $c$, and 4 internal parameters). Half 
of the atoms are in the plane $z=\frac{1}{4}c$, and the other half are in 
the $z=\frac{3}{4}c$ plane, as shown in Figure 1(b). Si atoms are four-fold
coordinated with N atoms in a slightly distorted tetrahedral configuration, 
and N atoms are threefold coordinated with Si atoms.
The six structural parameters 
$(a,c,x_{\rm Si},y_{\rm Si}, x_{\rm N}, y_{\rm N})$
optimized with GGA (7.667 \AA, 2.928 \AA, 0.1752, 0.7692, 0.3299, 0.031)
and LDA (7.585 \AA, 2.895 \AA, 0.1738, 0.7675, 0.3301, 0.0295) are in 
good agreement with previous calculations \cite{liubelkada}
and experimental values \cite{villars} of 
(7.608 \AA, 2.911 \AA, 0.1733, 0.7694, 0.3323, 0.0314).
The surface calculations were performed with the 
relaxed structural parameters using a slab geometry.
Depending on the surface, we used hexagonal and simple or base-centered
monoclinic supercells (Table I). For each surface, we considered up to 
4 different stoichiometric terminations to find the lowest-energy 
atomic configuration.
We performed several tests to assess the convergence of 
the surface energy with respect to number of layers (up to 7)
and the size of the vacuum region (up to 10 \AA).
Although the convergence was observed to 
be rapid [with the exception of the $(11\overline{2}0)$ surface],
we used 5-layer slabs [10-layer for
$(11\overline{2}0)$] with $8-10$ \AA\ vacuum. 

Experimentally, it is observed that the Si$_3$N$_4$ microstructure 
consists of elongated structures with hexagonal cross 
sections.
The preferential growth along the $c-$axis 
is rather rapid, while the growth of the $(10\overline{1}0)$
and $(11\overline{2}0)$ prismatic planes is 
reaction limited and very sensitive to the type of 
additives \cite{clarke-kleebe}. In order to understand the growth 
process of bare Si$_3$N$_4$ grains, we calculated the surface energies of 
five low-index Si$_3$N$_4$ surfaces (Table I).
The equilibrium shape of $\beta-$Si$_3$N$_4$ 
calculated from these energies by the
Wulff construction \cite{wulff} is shown in Figure 2.
The exposed faces consist of the \{$11\overline{2}0$\}, \{$10\overline{1}1$\}, 
\{$11\overline{2}1$\}, and \{$0001$\} families of planes making up 
$\sim$ 53 \%, 25 \%, 19 \%, and 3 \% of the total surface area, 
respectively. The calculated aspect ratio for bare surfaces
between the crystal length 
and its width is 1.4, which is in agreement with 
the measured aspect ratios near 2 for Si$_3$N$_4$
samples sintered without or only the minimum amount of
the additive.
We expect that in the presence of the rare-earth oxide additive, 
the energies of $(11\overline{2}0)$ or $(10\overline{1}0)$ will be 
lower, thereby increasing the calculated aspect ratio.
We also note that
our calculations predict an equilibrium 
shape, in which the base is {\em not} flat, but like a faceted dome
even when only two tilted families of planes,
\{$11\overline{2}1$\} and \{$10\overline{1}1$\}, are included. 
This is in agreement
with the experimental observation on macroscopic Si$_3$N$_4$ matrix-grains
embedded in the intergranular phase, where the basal planes look 
atomically rough \cite{kramer}.

\begin{table}
\caption{Supercell shape, parameters, and surface energies for the five
lowest-index $\beta-$Si$_3$N$_4$ surfaces. In the second column, the numbers
in parentheses are the surface unit cell dimensions; M (monoclinic) and BCM 
(base-centered monoclinic) refer to the lattice type of the supercell, and 
$\beta$ is the monoclinic angle. The energies for $(11\overline{2}0)$, 
$(11\overline{2}1)$, and $(10\overline{1}0)$ refer to the lowest-energy 
stoichiometric terminations, while the energy for $(10\overline{1}0)$ 
is that of the open-ring surface.}
\begin{ruledtabular}
\begin{tabular}{lcc}
Surface & Supercell & $E_{\rm surf}$ (J/m$^2$)\\ \hline
$(10\overline{1}0)$ & $(a,c)$, M, $\beta=60^{\circ}$ & 2.57\\
$(11\overline{2}0)$ & $(a\sqrt{3},c)$, M, $\beta=30^{\circ}$ & 1.95\\
(0001) & $(a,a)$, Hexagonal & 2.74\\
$(10\overline{1}1)$ & 
$(\sqrt{3a^2+4c^2},a)$, BCM, $\beta=66.2^{\circ}$ 
& 2.77\\
$(11\overline{2}1)$ & 
$(\sqrt{a^2+4c^2},a\sqrt{3})$, BCM, $\beta=52.6^{\circ}$ 
& 2.81\\
\end{tabular}
\end{ruledtabular}
\end{table}

\begin{figure}
\vspace{-1.0in}
\includegraphics[width=0.5\textwidth]{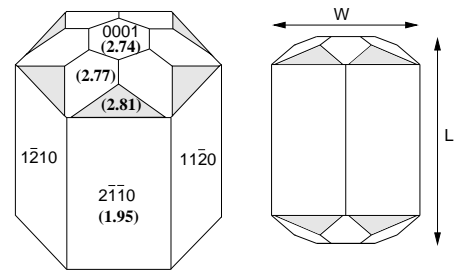}
\vspace{-1.5in}
\caption{Two views of the equilibrium shape of a $\beta$-Si$_3$N$_4$ crystal 
calculated using the Wulff construction. 
4 families of surfaces are exposed: \{$11\overline{2}0$\},
\{$0001$\},
\{$11\overline{2}1$\} (the shaded family of surfaces), and 
\{$10\overline{1}1$\} (the unshaded family of surfaces surrounding the 
$(0001)$ surface). The bold numbers in parentheses are the calculated
surface energies in J/m$^2$. The calculated $L/W$ aspect ratio is 1.4.}
\end{figure}

It is particularly interesting to notice that of the two
lowest-index prismatic planes rotated by 30$^{\circ}$ with respect to 
each other, it is the $(11\overline{2}0)$ surface
which is exposed in the Wulff construction 
and {\em not} the $(10\overline{1}0)$ surface with open hexagonal 
rings, which has been shown to exhibit an abrupt interface with the rare-earth
oxide additive in recent STEM experiments. The reason for the 
significantly lower surface energy of $(11\overline{2}0)$ (1.95 J/m$^2$)
compared to $(10\overline{1}0)$ (2.57 J/m$^2$) can be understood
from the nature of the atomic relaxations and the way the dangling bonds
of Si are saturated. On the ideal (unrelaxed) $(10\overline{1}0)$ surface,
there are one Si and one N atom, which have one dangling bond each.
These are shown by Si2 and N1 in Fig 3(a). When the surface
is relaxed, while N1 still remains with a dangling bond, 
Si2 undergoes a considerable displacement ($\sim$ 0.8 \AA) 
forming a new Si-Si bond with Si4 at 2.58 \AA, only 10 \% larger than 
the bulk Si-Si distance [Fig. 3(b)].  As such, the dangling bond
of Si2 is saturated, and the 
5-fold coordinated Si4 atom has two resonant
bonds, reminiscent of overcoordination of Si in negatively charged
$E-$center defect \cite{ogut2003}. The surface is still somewhat rough 
due to the presence of the Si1-N1 unit, which does not relax significantly.
On the ideal $(11\overline{2}0)$
surface, on the other hand, there are 2 Si and 2 N atoms with one 
dangling bond each, labeled by Si6, Si4, N5, and N3 in 
Fig. 3(c). When this surface is relaxed, the resulting atomic 
displacements for most of the atoms are very large. For example, 
the displacements of N7, Si5, and Si6 are 1.51 \AA, 1.49 \AA, and
1.33 \AA, respectively. Such large relaxations saturate 
all the dangling bonds except for N3, resulting in a rather smooth 
surface with 7-fold, 3-fold, and 4-fold Si-N rings [Fig. 3(d)]. The only
undercoordinated Si atom on the relaxed surface (Si5 with no 
dangling bonds on the ideal surface) exhibits an $sp^2-$type bonding
in an almost planar coordination with 3 N atoms. 

\begin{figure}
\vspace{-0.5in}
\includegraphics[width=0.6\textwidth]{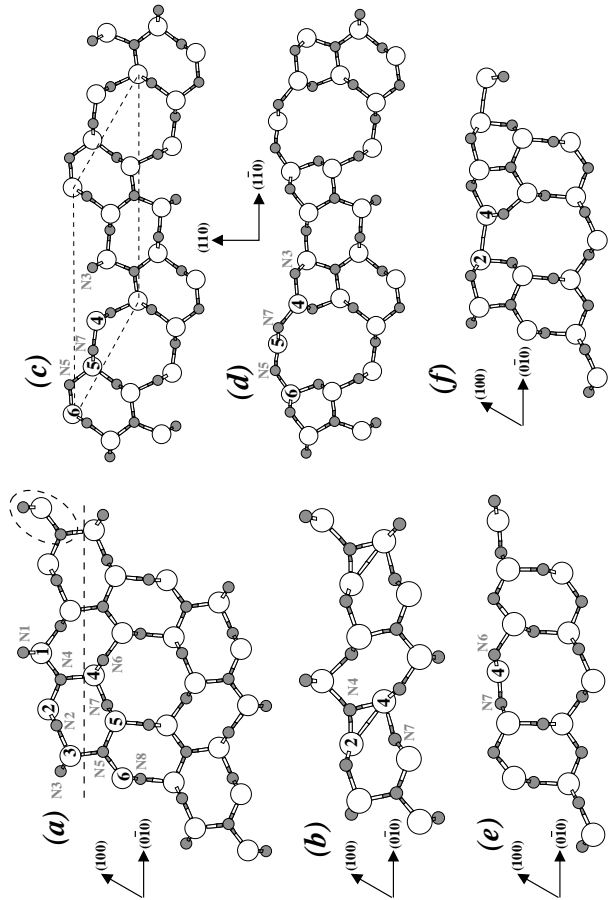}
\vspace{-0.5in}
\caption{$[0001]$ projected, double-unit-cell views of (a) 2-layer ideal 
$(10\overline{1}0)$ open-ring surface, (b) the top-layer of the relaxed
open-ring surface, (c) 2-layer ideal $(11\overline{2}0)$ surface, 
(d) 2-layer relaxed $(11\overline{2}0)$ surface, (e) top-layer of the relaxed
half-surface, and (f) relaxed structure of a
non-stoichiometric $(10\overline{1}0)$ surface with a missing SiN$_2$
unit (1.5 layers shown). In (a), the index 1, for example, inside a white
circle denotes the Si1 atom mentioned in the text.
If the atoms above the dashed line are removed, the ideal 
half-surface is obtained.
If the atoms inside the dashed ellipse
(N1, Si1, and N4) are removed, the ideal structure of the
non-stoichiometric surface (mentioned in the text) is obtained.
In (c), the dashed region corresponds to a single unit cell of the 
$(11\overline{2}0)$ surface.}
\end{figure}

We also investigated the relative structural stability of the 
$(10\overline{1}0)$ bare surface as a function of stoichiometry. 
The formation energy of a surface, $E_{\rm form}$, can be 
written as \cite{qian}
\begin{equation}
E_{\rm form}=
E_{\rm slab}^{\rm tot}-n_{\rm Si}\mu_{\rm Si}-n_{\rm N}\mu_{\rm N},
\end{equation}
where $E_{\rm slab}^{\rm tot}$ is the total energy of the slab, $n_i$ is the 
number of atoms of type $i$ in the slab, and $\mu_i$'s are the corresponding 
chemical potentials, which satisfy 
$3\mu_{\rm Si}+4\mu_{\rm N}=\mu_{\rm Si_3N_4,bulk}$. 
Figure 4 shows the calculated
energies as a function of the stoichiometry of the 
surface. All energies are
relative to the stoichiometric surface with open hexagonal
rings, which we will 
refer to as the ``open-ring surface". This 
reference configuration is quite stable in the full allowed region of
$\mu_{\rm N}$ with the exception of two other terminations.
The first one, which we will refer to as the ``half-surface",
is obtained by removing half the atoms from the open-ring surface
as shown by the dotted line in Fig. 3(a).
This surface, while stoichiometrically
terminated and quite smooth even in its ideal structure, 
does not have the open hexagonal rings, which are present
at the Si$_3$N$_4$ rare-earth oxide interfaces
studied in recent STEM experiments.
When relaxed, the half-surface
becomes 0.17 eV (GGA) and 0.1 eV (LDA) lower in energy compared to 
the open-ring surface.
The only significant relaxation occurs for the Si atom (Si4)
with the dangling 
bond which moves inward by $\sim 0.64$ \AA.
As a result, the Si atom
exhibits a $sp^2-$type bonding in a
nearly planar coordination with 3 N atoms [Fig. 3(e)].
One possibility for the discrepancy between the theoretical prediction 
of the half-surface having a lower energy than the experimentally 
observed open-ring surface
at the rare-earth oxide interface 
is that theory incorrectly predicts the lowest energy
termination of the bare $(10\overline{1}0)$ Si$_3$N$_4$ surface. 
This could be verified by studying the bare
surface with low-energy electron diffraction experiments.
A more likely reason
for the discrepancy, which would also explain the prediction of the 
$(11\overline{2}0)$ surface as the lower-energy prismatic plane
contrary to experimental observations, is that 
the rare-earth oxide additive changes
the relative stability of different Si$_3$N$_4$ surfaces and 
different terminations of the same-index surfaces.

\begin{figure}
\vspace{-0.8in}
\includegraphics[width=0.5\textwidth]{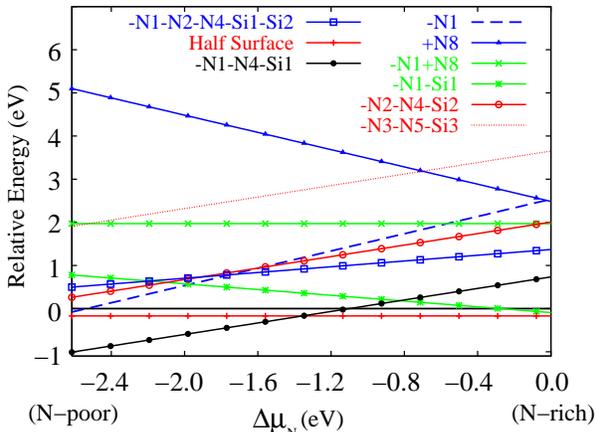}
\vspace{-1.5in}
\caption{(Color) The formation energies of various terminations of the $(10\overline{1}0)$
surface relative to the open-ring surface
as a function of stoichiometry
$\Delta\mu=\mu_{\rm N}-\frac{1}{2}\mu_{\rm N_2}$. $(+)$ and $(-)$ signs
correspond to taking out and adding the particular atom, respectively,
from the open-ring
surface shown in Fig. 3(a) with all the atom indices.}
\end{figure}

The second termination, which results in a lower energy surface
under N-poor conditions, is obtained by removing a SiN$_2$ unit
(N1, N4, Si1) from the open-ring surface, as shown by the dashed 
ellipse in Fig. 3(a). In its ideal structure, this surface has 2 Si 
atoms, Si4 and Si2, with 1 and 2 dangling bonds, respectively, along
with a N atom (N3) with one
dangling bond. In spite of a large number of broken bonds, when 
the surface is relaxed, large atomic displacements of Si2 (1.29 \AA)
and Si4 (1.27 \AA) saturate all Si dangling bonds
(via the formation of a new Si-Si bond at 2.5 \AA),
which reduces the surface energy considerably.
The relaxed surface is smooth and has 7-fold and 
3-fold rings, similar to those observed on the $(11\overline{2}0)$ 
surface [Fig. 3(f)].
It is interesting to note that under extreme N-poor conditions, the 
energy of this surface is 1.85 J/m$^2$, which is lower than that of
$(11\overline{2}0)$, providing a possible explanation for the 
dominant observation of $(10\overline{1}0)$ surfaces in the STEM 
experiments.

In summary, we have presented results from {\em 
ab initio} calculations on the atomic and electronic structures 
of bare $\beta-$Si$_3$N$_4$ surfaces with emphasis
on the prismatic plane $(10\overline{1}0)$ surface 
observed in recent STEM experiments. 
The equilibrium shape of a macroscopic Si$_3$N$_4$ crystal is found to  
have a hexagonal cross section, a faceted dome-like base, and 
an aspect ratio of 1.4, 
in agreement with experimental observations.
We find large distortions on the prismatic planes
driven primarily by the tendency 
of Si atoms to saturate their dangling bonds and 
achieve either resonant-bond, $sp^3-$, or $sp^2-$bonded configurations.
The stoichiometric $(11\overline{2}0)$ surface, 
the $(10\overline{1}0)$ ``half-surface",
and a non-stoichiometric $(10\overline{1}0)$ surface 
obtained by removing a SiN$_2$ unit are predicted 
to have lower energies than the
open-ring $(10\overline{1}0)$ surface.
In light of the consistent experimental
observations of the open-ring $(10\overline{1}0)$ 
surface at the interface,
the present results obtained with state-of-the-art
{\em ab initio} techniques strongly indicate that (i)
the rare-earth oxide additive changes the relative stability of 
$\beta-$Si$_3$N$_4$ surfaces, and (ii) non-stoichiometry (especially 
resulting from N-poor conditions) should play an important role in 
determining the termination of the Si$_3$N$_4$ matrix grains. 
We expect that future {\em ab initio} studies of the interface will
not only focus on the bonding nature of the additives to the open-ring
$(10\overline{1}0)$ surface, but also address why the proposed mechanisms
do not favor promotion of the $(11\overline{2}0)$ termination, which has a
lower bare surface energy.
This work was supported by DOE under grants No. DE-FG02-03ER15488 
(JCI and S\"{O}), DE-AC05-03ER46057 (AZ and NDB),
and DE-AC02-05CH11231 (ROR), and by the
ACS Petroleum Research Fund under grant No. 40028-AC5M (HI).

\end{document}